\documentclass[prl,aps,twocolumn,superscriptaddress,showpacs]{revtex4-1}
\usepackage{amsmath,amssymb,graphicx}
\usepackage{epsfig}
\usepackage{graphicx}
\usepackage{textcomp}

\begin{document}

\title{Extreme Current Fluctuations in Lattice Gases: Beyond Nonequilibrium
Steady States}

\author{Baruch Meerson}
\email{meerson@mail.huji.ac.il}
\affiliation{Racah Institute of Physics, Hebrew University of
Jerusalem, Jerusalem 91904, Israel}
\author{Pavel V. Sasorov}
\email{pavel.sasorov@gmail.com}
\affiliation{Keldysh Institute of Applied Mathematics, Moscow 125047, Russia}



\begin{abstract}
We use the macroscopic fluctuation theory (MFT) to study large current fluctuations
in non-stationary diffusive lattice gases.  We identify two universality classes
of these fluctuations which we call elliptic and hyperbolic. They emerge in the limit when the deterministic mass flux is small compared to the mass flux due to the shot noise. The two
classes are determined by the sign of compressibility of \emph{effective fluid},  obtained by mapping the MFT
into an inviscid hydrodynamics. An example of the elliptic class is the Symmetric Simple Exclusion Process where,
for some initial conditions, we can solve the effective hydrodynamics exactly. This
leads to a super-Gaussian extreme current statistics
conjectured by Derrida and Gerschenfeld (2009) and yields the optimal path of the system.
For models of the hyperbolic class the deterministic mass flux
cannot be neglected, leading to a different extreme current statistics.
\end{abstract}
\maketitle

Large fluctuations of currents of matter or energy in systems away from thermodynamic equilibrium are at the forefront of statistical physics. Over the last decade a  major progress has been achieved in the study of large fluctuations
of the density profile and of the current in \emph{nonequilibrium steady states} (NESS) of stochastic diffusive lattice gases
driven from the
boundaries, see \cite{D07} and references therein.  Diffusive lattice gases \cite{KL99,Spohn,L99} constitute a broad family of
simple transport models which capture
different aspects of transport in extended many-body systems. One extensively studied model
is the Symmetric Simple
Exclusion Process (SSEP)
\cite{Spohn,KL99,L99,SZ95,D98,S00,BE07,KRB10},
where each particle can randomly hop
to a neighboring lattice site
if that site is empty. If it is occupied by another particle, the move is disallowed. Applications of this model range from full counting statistics of mesoscopic
conductors \cite{Levitov,Buttiker,Jordan1,Jordan2,D07} to a host of transport problems in
materials science, cell biology and biophysics \cite{Chou}.

The large deviation functionals \cite{NESS} of the density and the current of NESS of diffusive lattice gases
exhibit qualitatively new features compared to the free energy of
equilibrium states  \cite{D07,Jona,Hurtado,Kafri},
and these discoveries have attracted great interest.  \emph{Non-stationary} fluctuations of diffusive lattice gases are
still poorly understood \cite{DG2009a,DG2009b,varadhan,KM_var,van,void}, and they will be in
the focus of our attention here. Following Refs. \cite{DG2009a,DG2009b,varadhan,KM_var}, we will consider a diffusive
lattice gas on an infinite line, and study fluctuations
of integrated current $J$ through the origin $x=0$ during a fixed time $T$,
when starting from a deterministic step-like density profile
\begin{equation}
\label{step}
n(x,t=0) = n_-\theta(-x) + n_+ \theta(x),
\end{equation}
where $\theta(x)$ is the Heaviside step function. (Here and in the following by density we mean the number of particles per lattice site.)  In deterministic limit, the large-scale behavior of diffusive lattice gases is described by
the diffusion equation $\partial_t n = \partial_x \!\left[D(n) \,\partial_x n\right]$. Solving it with the initial condition \eqref{step}, one can compute the \emph{average} integrated current
at time $T$: $\langle J(T)\rangle = \int_0^\infty dx\, [n(x,T)-n_+]$.
The actual current $J$ fluctuates.  
At large scales these
fluctuations can be described by 
the Langevin equation
\begin{equation}
\label{Lang}
     \partial_t n = \partial_{x} [D(n)\, \partial_x n] +\partial_x \left[\sqrt{\sigma(n)} \,\eta(x,t)\right],
\end{equation}
where $\eta(x,t)$ is a zero-average Gaussian noise, delta-correlated both in space and in time \cite{Spohn,KL99}.
As one can see, a fluctuating lattice gas is fully characterized by $D(n)$ and the coefficient $\sigma(n)$, which comes from the shot noise and is equal to twice the mobility of the gas \cite{Spohn,Jona}.

Starting from Eq.~(\ref{Lang}), one can arrive at \emph{macroscopic fluctuation theory} (MFT), which employs $1/\sqrt{N}$ ($N$ is the typical number of particles in the relevant region of space) as a small parameter, and is appropriate for dealing with large deviations. The MFT was originally developed for the NESS \cite{Bertini} and more recently extended to non-stationary settings, such as the step-like initial density profile \cite{DG2009b,KM_var}. The MFT can be formulated as a classical Hamiltonian
field theory \cite{Bertini,DG2009b,Tailleur,KM_var}, and we will adopt this formulation here.

The MFT formulation of the problem of statistics of integrated current was obtained in Ref. \cite{DG2009b}.
Until now the problem has defied analytic solution except (i) for
non-interacting random walkers (RWs) \cite{DG2009b}, and (ii) for small fluctuations around the
mean \cite{varadhan,KM_var}.
For the RWs, the $J \to \infty$
asymptote of the current probability density ${\cal P}(J,T)$ is super-Gaussian in $J$ \cite{DG2009b}:
\begin{equation}\label{PRW}
\ln {\cal P}(J\to \infty,T)\simeq - \frac{J^3}{12 n_{-}^2 T}.
\end{equation}
Derrida and Gerschenfeld \cite{DG2009b} (DG) conjectured that the $\sim J^3/T$ decay holds for a whole class of
\emph{interacting} gases, and proved their conjecture for $D(n)=1$
and $\sigma(n)\leq n+\text{const}$ for $0\leq n\leq R$, and $\sigma(n)=0$ otherwise \cite{varadhan2}.

This Letter reports a major progress in the analysis of extreme current fluctuations. Here is an outline. A natural first step in the analysis of
unusually large currents is to neglect, in the MFT equations, the deterministic mass flux compared to the mass
flux due to the shot noise. The noise-dominated MFT equations can then be mapped into an effective inviscid hydrodynamics. This
mapping uncovers two universality
classes (which we call elliptic and hyperbolic) of the diffusive lattice gases with respect to the extreme current statistics. These are determined by the sign of
$\sigma^{\prime\prime}(n)$. For
the elliptic class  $\sigma^{\prime\prime}(n)<0$ for all relevant $n$. Here  the DG
conjecture holds, as
\begin{equation}\label{Pgeneric}
\ln {\cal P}(J\to \infty,T)\simeq -\frac{f(n_{-},n_+) J^3}{T}.
\end{equation}
Furthermore, for the SSEP, with $\sigma(n)=2n(1-n)$ \cite{Spohn,KL99}, the effective hydrodynamics can be solved exactly. The solution yields a systematic way of calculating the function $f(n_-,n_+)$ and gives the optimal path of the system, responsible for the specified current.

The hyperbolic case, when $\sigma^{\prime\prime}(n)>0$,
is more complicated. Here a singularity, present in  the noise-dominated equations, has to
be regularized by diffusion. The resulting $\ln P$ 
differs from Eq.~(\ref{Pgeneric}) \cite{MSKMP}.
Now we expose our results
in some detail.

\textsc{MFT.}
A specified
current
is described by the equation
\begin{equation}
\label{current0}
\int_0^\infty dx\, [n(x,1)-n_{+}] =J/\sqrt{T}\equiv j.
\end{equation}
Here and in the following $t$ and $x$ are rescaled by  $T$ and $\sqrt{T}$, respectively.  The optimal path $q(x,t)$ in the space of $\{n(x,t)\}$
obeys the equations
\begin{eqnarray}
\partial_t q &=& \partial_x \left[D(q)\, \partial_x q\right]
-  \partial_x \left[\sigma(q)\, \partial_x p\right],\label{q:eqfull}\\
\partial_t p &=& - D(q) \partial_{x}^2 p
- \frac{1}{2}\sigma^{\prime}(q)\!\left(\partial_x p\right)^2,\label{p:eqfull}
\end{eqnarray}
where $p(x,t)$ is the ``momentum" field  conjugate to $q$ \cite{DG2009b}. These are Hamilton equations,
with the Hamiltonian 
$H=\int_{-\infty}^\infty dx\,\mathcal{H}$, where $\mathcal{H}(q,p) = -D(q)\, \partial_x q\, \partial_x p
+(1/2) \sigma(q)\!\left(\partial_x p\right)^2$. Once $q(x,t)$ and $p(x,t)$ are known, the action $\int\int dt dx \left(p\,\partial_t q - \mathcal{H}\right)$ can be written as \cite{Tailleur,DG2009b,KM_var}
\begin{eqnarray}
  s  = \frac{1}{2}\int_0^1 dt \int_{-\infty}^\infty dx \,\sigma(q) (\partial_x p)^2. \label{action0}
\end{eqnarray}
The boundary condition for $q(x,0)$ is given by $n(x,t=0)$ from Eq.~\eqref{step}. By varying $q(x,1)$ to minimize the action under the constraint~(\ref{current0}), DG \cite{DG2009b} obtained
the second boundary condition:
\begin{equation}
\label{p_step}
p(x,t=1) = \lambda \,\theta(x),
\end{equation}
where the Lagrange multiplier
$\lambda=\lambda(j,n_{-},n_{+})>0$ is set by Eq.~(\ref{current0}).  Once $s$ is
found, ${\cal P}(J,T)$ is given by $\ln {\cal P}(J,T) \simeq -\sqrt{T} \,s$ \cite{DG2009b,KM_var}.

\textsc{Inviscid limit.}  When $j\to \infty$, it seems natural to  neglect the deterministic diffusion
terms in Eqs. (\ref{q:eqfull}) and (\ref{p:eqfull}).
This yields the first-order
Hamiltonian equations
\begin{equation}
\partial_t q \!=\! -  \partial_x\left[\sigma(q)\, \partial_x p\right],\,\text{(a)}\;\;
\partial_t p \!= \!- \frac{1}{2}\sigma^{\prime}(q)\!\left(\partial_x p\right)^2,\;\text{(b)} \label{qp:eq}
\end{equation}
stemming from the \emph{noise-dominated} Hamiltonian
\begin{equation}
\label{Ham}
H_0=\int_{-\infty}^{\infty} dx \,\rho (q,p),\;\;\text{where}\;\;\rho = \frac{1}{2}\sigma(q)\!\left(\partial_x p\right)^2.
\end{equation}
We will call this theory inviscid. Comparing Eqs.~(\ref{action0}) and (\ref{Ham}) and using the fact that $H_0$
is now a constant of motion, we can rewrite Eq.~(\ref{action0}) as $s=\int_0^1 dt\, H_0=H_0$. The inviscid MFT equations
are invariant under the transformation $x/\sqrt{\lambda} \to x$
and $p/\lambda \to p$. Under this transformation $s$ becomes $\lambda^{3/2} s_1$,
where $s_1$ is the action obtained with the condition
(\ref{p_step}) replaced by $p(x,1)=\theta(x)$. In its turn, Eq.~(\ref{current0}) becomes
$\int_0^{\infty}dx\,[q(x,1)-n_+]=j/\sqrt{\lambda}$, therefore $j=\sqrt{\lambda} j_1$,
where $j_1$ is the integrated current obtained  for $p(x,1)=\theta(x)$.
Putting everything together, we arrive at Eq.~(\ref{Pgeneric}) with $f(n_-,n_+)=s_1/j_1^3$.
Therefore, the DG conjecture is a natural consequence of the inviscid MFT.
It remains to be seen, however, whether the inviscid problem is well defined.

\begin{figure}
\includegraphics[width=1.8 in,clip=]{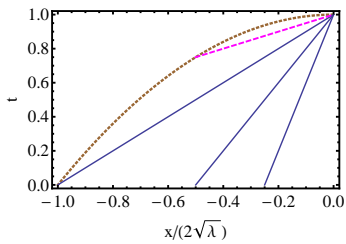}
\caption{(Color online) The characteristics of Eq.~(\ref{qp:eq})a for the RWs. Here $\sigma(n)=2n$,
and $\partial_x p$ is given by Eq.~(\ref{vRW}). The solid lines: characteristics crossing
the boundary $t=0$. The dashed line: a characteristic crossing the ``velocity" shock at
$x=x_s(t)=-2\sqrt{\lambda (1-t)}$. The velocity shock position is denoted by the dotted line.}
\label{characteristics}
\end{figure}

\textsc{Independent Random Walkers (RWs).} The first check is the case of RWs, where $\sigma(n)=2n$ \cite{Spohn,KL99}.
Here Eq.~(\ref{qp:eq})b
reduces to the Hopf equation $\partial_t p+(\partial_x p)^2=0$.
Its solution, obeying the condition~(\ref{p_step}), yields \cite{Whitham}
\begin{equation}\label{vRW}
    \partial_x p(x,t) = \left\{\begin{array}{ll}
-\frac{x}{2(1-t)}, & \mbox{$-2\sqrt{\lambda (1-t)}<x<0$}, \\
0  & \text{elsewhere}
\end{array}
\right.
\end{equation}
(we do not use here the rescaling with $\lambda$). Equation (\ref{vRW}) includes a
shock at $x_s(t)=-2\sqrt{\lambda (1-t)}$.  Now, Eq.~(\ref{qp:eq})a is a continuity
equation for the density $q(x,t)$ with a
known velocity field $2\partial_x p$. The characteristics of this equation are
$x=C (1-t)$, where $C=\text{const}$.
Consider the region of the $x,t$ plane
where $\partial_x p\neq 0$, see Fig. \ref{characteristics}.  The characteristics
with $-2\sqrt{\lambda}\leq C \leq 0$ cross the boundary $t=0$,
where
$q(x,t=0)=n_{-}$. This yields a simple solution, $q(x,t)=n_{-}(1-t)^{-1}$, to the right of the
characteristic $x=-2\sqrt{\lambda}(1-t)$.

The characteristics with $C<-2\sqrt{\lambda}$ cross the velocity
shock at $x=x_s(t)$. To ensure mass conservation at the shock (where the velocity drops
from a positive value to zero), $q$ at the shock must vanish. As a result,
$q(x,t)=0$ on the interval $-2\sqrt{\lambda}<x<-2\sqrt{\lambda}(1-t)$ which expands in time. To
summarize,
\begin{equation}
q(x,t)=\left\{\begin{array}{llll}
n_{-}, & \mbox{$x<-2\sqrt{\lambda}$}, \nonumber \\
0, & \mbox{$-2\sqrt{\lambda}<x<-2\sqrt{\lambda} (1-t)$}, \nonumber \\
n_{-}(1-t)^{-1}, & \mbox{$-2\sqrt{\lambda} (1-t)<x<0$}, \nonumber \\
n_{+},  & \mbox{$x>0$.}
\end{array}
\right.
\end{equation}
Importantly, at $t=0$  the flow already includes a point-like void, $q=0$, at the
point $x=-2\sqrt{\lambda}$ \cite{regularization}. At $t>0$ the void expands. Simultaneously,
a constant mass of gas, equal to $2n_{-}\sqrt{\lambda}$, gets squeezed on the interval
$-2\sqrt{\lambda} (1-t)<x<0$ which shrinks with time. At $t=1$ all this mass collapses into
the point $x=0$, leaving behind a void at $-2\sqrt{\lambda}<x<0$.  The collapsed mass
is
equal to the (rescaled) integrated current $j$, see Eq.~(\ref{current0}), and we
obtain $\lambda=j^2/(4n_{-}^2)$. The rescaled action (\ref{action0}) is
\begin{eqnarray}
\label{sinviscid}
  s&=& \int_0^1 dt \int_{-2\sqrt{\lambda} (1-t)}^{0} dx \,\, \frac{n_{-}}{1-t} \,\, \frac{x^2}{4 (1-t)^2}\nonumber \\
  &=& \frac{2n_{-}\lambda^{3/2}}{3} = \frac{j^3}{12 n_{-}^2},
\end{eqnarray}
which, in view of the relation $\ln {\cal P}(J,T) \simeq -\sqrt{T} \,s$, yields Eq. (\ref{PRW}). As we see,
for the RWs, the inviscid MFT does yield the correct leading-order result for the extreme current
statistics and for the optimal path of the system.

\textsc{Effective hydrodynamics and two universality classes.}  Now let us consider some
general properties of the inviscid theory as applied to \emph{interacting} particles. A
key observation is that the Hamiltonian density $\rho$ of the inviscid MFT, see Eq. ~(\ref{Ham}),
is conserved \emph{locally}, as
it evolves according to the continuity equation
\begin{equation}\label{inviscid1}
    \partial_t \rho +\partial_x (\rho V) = 0
\end{equation}
with the effective velocity $V=\sigma^{\prime}(q) \partial_x p$. A direct calculation shows that
$V$ obeys
\begin{equation}\label{inviscid2}
\partial_t V+V\partial_x V= -\sigma^{\prime\prime} (q) \partial_x \rho,
\end{equation}
where it is assumed that $\sigma^{\prime\prime}(q)$ is expressed via $\rho$ and $V$.
Equations (\ref{inviscid1}) and (\ref{inviscid2}) describe an inviscid hydrodynamic flow of an effective
fluid.  The general character of this
flow  --  elliptic or hyperbolic -- is determined by whether $\sigma^{\prime\prime}(q)$ is negative or positive,
respectively. The analogy with hydrodynamics becomes complete when $\sigma(\rho)$ is a quadratic polynomial,
and so the right side of Eq.~(\ref{inviscid2}) can be written as $-(1/\rho)\, \partial_x P(\rho)$. This is what happens
for the SSEP, where the effective fluid pressure is $P(\rho)=-2 \rho^2<0$, exemplifying
elliptic flow. In an \emph{initial-value} problem such a fluid would be intrinsically
unstable. Moreover,
when starting from generic smooth fields $\rho$ and $V$ at $t=0$, a finite-time singularity develops, see Ref. \cite{Trubnikov}
for a detailed review. In our boundary-value problem, $\rho$ and $V$ must blow up, in view of Eq.~(\ref{p_step}), at time $t=1$ at $x=0$. However, they are bounded and smooth at earlier times, $0\leq t<1$, in the whole region where $\rho(x,t)>0$. Therefore, the inviscid MFT is well defined for elliptic flows, and Eq.~(\ref{Pgeneric}) holds \cite{total mass}.

Models where $\sigma^{\prime\prime}(n)>0$
exhibit a hyperbolic flow. 
In view of the boundary condition~(\ref{p_step})
$\rho$ and $V$ must diverge at $t=1$ and $x=0$.  In a hyperbolic
flow this implies that the singularities $\rho=\infty$ and $V=\infty$ are present
\emph{at all times} $0\leq t\leq 1$.
Here one needs to return to the full MFT equations (\ref{q:eqfull})
and (\ref{p:eqfull}), where the singularities are regularized by diffusion. This leads
to a different extreme current statistics \cite{MSKMP}.

The RWs, with $\sigma(n)=2n$, belong to the marginal class $\sigma^{\prime\prime}=0$ where the effective fluid
pressure vanishes, leading to the Hopf equation as we already observed.

\textsc{The SSEP.}
Fortunately, Eqs.~(\ref{inviscid1}) and (\ref{inviscid2}) become linear upon the hodograph
transformation, where $\rho$ and $V$ are treated as the independent variables, and $t(\rho,V)$
and $x(\rho,V)$ as the dependent ones \cite{LLfluidmech}. For the SSEP we obtain
the elliptic linear second-order equation
\begin{equation}\label{thodograph}
 \rho^{-1} \partial_{\rho}\left(\rho^2 \partial_{\rho}t\right)+4\partial_{V}^2 t=0.
\end{equation}
Once $t(\rho,V)$ is found, $x(\rho,V)$ can be determined from any of the relations
\begin{equation}\label{xhodograph}
\partial_V x=V\partial_V t-\rho \partial_{\rho} t,\;\;\; \partial_{\rho} x=V \partial_{\rho} t +4 \partial_V t.
\end{equation}
In the new variables  $X=V/2$ and $Y=2\rho^{1/2}$
Eq.~(\ref{thodograph}) becomes
\begin{equation}\label{preLaplace}
\partial_{X}^2 t+\partial_{Y}^2 t+(3/Y) \,\partial_Y t
=0.
\end{equation}
As we will see shortly, the boundary conditions for $q$ and $p$ at $t=0$ and $1$, respectively,
define a Dirichlet problem for $t(X,Y)$.  Moreover,  Eq.~(\ref{preLaplace}) can be transformed
into the Laplace's equation in an extended space \cite{Trubnikov}, which
opens the way to exact analytic solution. The full solution also includes non-hodograph regions: (i) static regions where
$q=n_{-}$ or $n_{+}$ and $\partial_x p =0$,
(ii) a void, $q=0$,  and (iii)
a close-packed cluster, $q=1$. The dynamics of $p$ in the void and cluster regions is described by the Hopf equation
$\partial_t p \pm (\partial_x p)^2=0$. As $\sigma(0)=\sigma(1)=0$, the non-hodograph regions
do not contribute to the action (\ref{action0}).  Importantly, (point-like) void and
cluster are already present at $t=0$; they expand at $t>0$.

\begin{figure}
\includegraphics[scale=0.45]{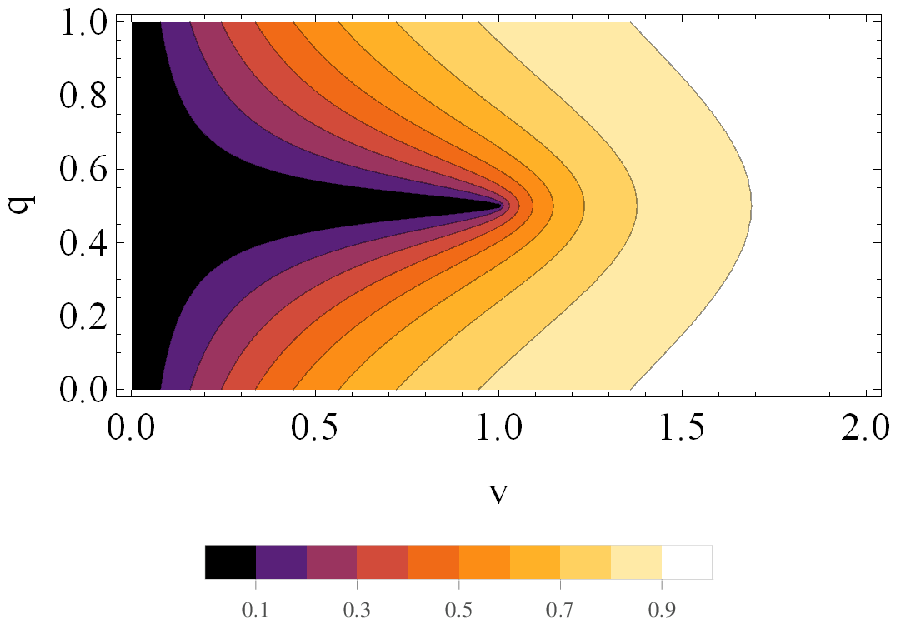}
\includegraphics[scale=0.45]{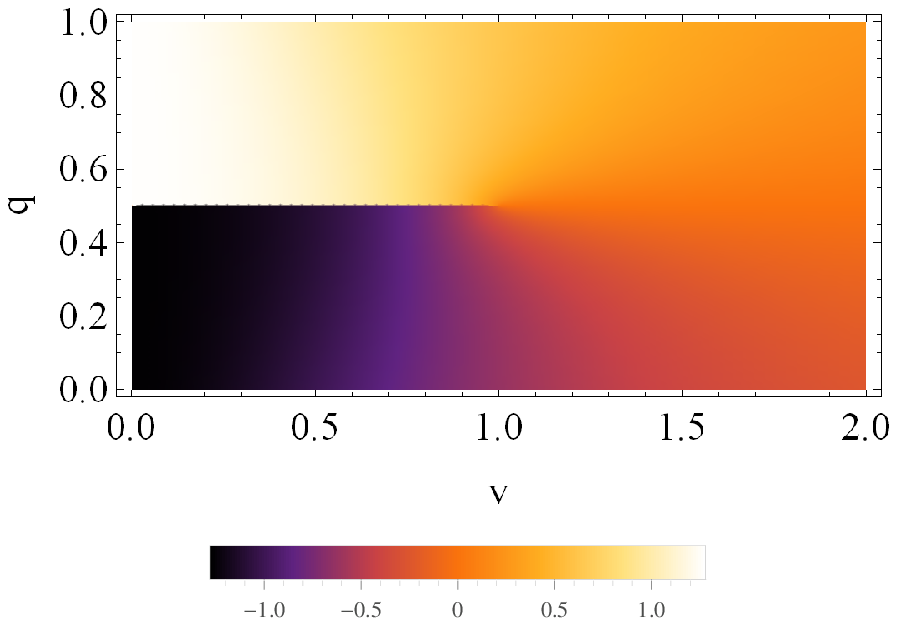}
\caption{The density plots of the hodograph solutions (\ref{telem}) for $t(q,v)$ (left) and (\ref{xelem}) for $x(q,v)$ (right),
for the SSEP with $n_-=n_+=1/2$. $x$ is measured in units of $\pi j/2$.}
\label{hsolution}
\end{figure}

Here
we only consider the simple case of a flat initial density profile with $q(x,t=0)=1/2$. In this case
the solution
is symmetric, $q(-x,t)=1-q(x,t)$ and $\partial_x p(-x,t)=\partial_x p(x,t)$ and, remarkably,
can be
obtained in elementary functions.  We start from Eq.~(\ref{preLaplace}) which should be solved
in the upper half-plane $|X|<\infty$, $0\leq Y<\infty$. To match the non-hodograph part of the
full solution, the hodograph solution must be regular at $Y=0$. The second boundary
condition comes from $t=0$. The value of $\partial_x p(x,t=0)$ changes, as a function of $x$,
from $0$ to an a priori unknown finite maximum value $v_0>0$. Exploiting the invariance of
the inviscid MFT equations under the transformation
$x/\sqrt{\lambda}\to x$ and $p/\lambda \to p$, we can first solve the problem for
$v_0=1$ and then restore the $\lambda$-scalings in the final solution. By virtue of the conditions
$q(x,t=0)=1/2$ and $0\leq\partial_x p\,(x,t=0)\leq 1$, $t$ must vanish on the segment $X=0,\,0\leq Y\leq 1$.
The last boundary condition comes from Eq.~(\ref{p_step}) at $t=1$. As $\partial_x p\,(x,t=1)$ is a delta-function,
$t$ must approach $1$ as $|X|$ or $Y$ go to infinity.
The solution of this Dirichlet problem can be
obtained in the elliptic coordinates $s, r$: $X=sr$, $Y=[(1+s^2)(1-r^2)]^{1/2},\;\; s\geq 0,\;\; |r|\leq 1$.
After some algebra, Eq.~(\ref{preLaplace}) becomes
\begin{equation}\label{telliptic}
\partial_s \left[(1+s^2)^{2}\, \partial_s t\right]
+\frac{1+s^2}{1-r^2}\,\partial_r \left[(1-r^2)^{2}\, \partial_r t\right]=0\, .
\end{equation}
The new boundary conditions, $t(s=0,|r|\leq 1)=0$ and $t(s\to \infty)=1$, are independent of $r$,
and so is the solution: $\pi t(s)/2=s (1+s^2)^{-1}+\arctan s$.
In the original variables $q$ and $v=\partial_xp$, we obtain
\begin{equation}
\label{telem}
  \frac{\pi}{2}\,t(q,v) = \frac{\sqrt{2(v^2+{\cal R}-1)}}{v^2+{\cal R}+1}
    +\arctan \sqrt{\frac{{v^2+\cal R}-1}{2}},
\end{equation}
where $\mathcal{R}^2(q,v)= (v^2-1)^2+4v^2 (2q-1)^2$. Then Eqs.~(\ref{xhodograph}) yield
\begin{equation}
\label{xelem}
x(q,v)=
 -\frac{\left[v^2-{\cal R} -8q(1-q)+1\right]\sqrt{v^2+{\cal R}-1}}{2\sqrt{2}\,\pi\, q (1-q) (2q-1) v}.
\end{equation}
Functions $t(q,v)$ and $x(q,v)$ are analytic in the whole hodograph region $0\leq q \leq 1,\, 0\leq v < \infty$ except at the branch cut at $q=1/2,\,0\leq v \leq 1$, see Fig.~\ref{hsolution}.

The hodograph asymptotics
at $v\to \infty$ are $t\simeq 1-4/(3\pi v^3)$ and $x \simeq 4 (2 q-1)/(\pi v^2)$. In physical variables,
we obtain
self-similar asymptotics at $t \to 1$: $2q(x,t)-1=x/\ell(t)$ and $v(x,t)=(4/3\pi)^{1/3}(1-t)^{-1/3}$,
where $|x|\leq \ell(t)$ and $\ell(t)=3(4/3\pi)^{1/3}(1-t)^{2/3}$. Using  Eq.~(\ref{action0}) or (\ref{Ham}),
we obtain
$s=H_0=\int_{-\ell(t)}^{\ell(t)} dx \,q(1-q) v^2|_{t\to 1}=4/(3\pi)$.

\begin{figure}
\includegraphics[width=1.7 in,clip=]{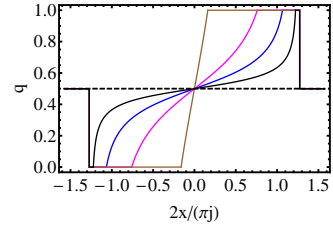}
\caption{(Color online) The optimal path found analytically: $q$ versus $x$ at $t=0$ (dashed line), $0.25$, $0.5$, $0.75$ and $0.98$
for the SSEP with $n_-=n_+=1/2$.}
\label{solution}
\end{figure}

In its turn, the $q\to 1/2$  asymptotic of Eq.~(\ref{xelem}) corresponds to $t\to 0$, and we obtain
$v(x,t=0)= (1-x^2/x_0^2)^{1/2}$,
for $|x|\leq x_0$, and $0$ otherwise, where $x_0=4/\pi$.  Notably,  $x=-x_0$ and $x=x_0$ are
the positions of the point-like void and point-like cluster, respectively, at $t=0$.
At $t>0$ the void and cluster expand, see Fig. \ref{solution}, and by $t=1$ all of
the material from the interval $-x_0<x<0$
is transferred to the interval $0<x<x_0$. Therefore, the integrated current is
$j_0=(1/2) \times x_0 =2/\pi$, whereas
$\lambda_0=\int_{-x_0}^{x_0} dx\, v(x,t=0) = 2$. Restoring the $\lambda$- and $j$-scalings,
we obtain $\lambda=\pi^2 j^2/2$ and $s=\pi^2 j^3/6$, which leads to Eq.~(\ref{Pgeneric}) with
$f(1/2,1/2)=\pi^2/6$. (See Supplemental Material below
for an alternative derivation of this result.) This completes the evaluation of $\ln {\cal P}(J,T)$
and justifies the inviscid MFT for the SSEP.

The inviscid MFT can be extended to other non-stationary settings. One of them is the noise-driven void formation, at a specified time $T$, in an initially uniform gas \cite{void},
in the limit of $L\gg \sqrt{T}$, where $L$ is the characteristic void size. As it is evident from
Fig. \ref{solution}, the void formation problem is closely related to the extreme current problem. It also extends to higher dimensions.

We thank Arkady Vilenkin, whose numerical insight directed us in the initial stage of this work,
and P.L. Krapivsky for useful discussions. B.M.
was supported by the Israel
Science Foundation (Grant No. 408/08) and by the US-Israel Binational Science Foundation (Grant No. 2008075).
P.V.S. was supported by the Russian Foundation for Basic Research, grant No 13-01-00314.

\appendix

\section{Supplemental Material: Alternative derivation of the result $f(1/2,1/2)=\pi^2/6$ for the SSEP}
\label{aa}
The same result $f(1/2,1/2)=\pi^2/6$  for the SSEP can be extracted from the expressions obtained by Derrida and Gerschenfeld (DG) \cite{DG2009a,DG2009b}. DG employed the moment generating function of $J$:
\begin{equation}
\label{GF}
\left\langle e^{\lambda J} \right\rangle = \sum_{J\geq 0} e^{\lambda J} P(J).
\end{equation}
For diffusive lattice gases with a step-like initial condition one has, at long times,
$\ln P(J) \simeq -\sqrt{T}\, s(j,n_-,n_+)$, where $j =J/\sqrt{T}$, see the main text.
Therefore, 
\begin{equation}
\label{asymp}
\left\langle e^{\lambda J} \right\rangle  \sim \int_0^{\infty} dj\, e^{\sqrt{T}\left[\lambda j-s(j,n_-,n_+)\right]} \sim
e^{\sqrt{T}\,\mu(\lambda,n_-,n_+)},
\end{equation}
where $\mu(\lambda,n_-,n_+)= \max_{j}\, [\lambda j-s(j,n_-,n_+)]$. That is, $\mu(\lambda)$ is the Legendre transformation of $s(j)$.  In reverse, $s(j)$ is the Legendre transformation of $\mu(\lambda)$:
\begin{equation}\label{reverse}
    s(j,n_-,n_+)=\max_{\lambda}\, [j \lambda-\mu(\lambda,n_-,n_+)].
\end{equation}

Using the microscopic model of SSEP, DG \cite{DG2009a} calculated $\mu (\lambda)$ in the \emph{annealed} setting, that is when the initial densities at $x<0$ and $x>0$ are allowed to fluctuate around their respective mean values $n_-$ and $n_+$, and one averages the result over these fluctuations. DG obtained
\begin{equation}
\label{annealed}
\mu_{\text{annealed}} (\lambda,n_-,n_+)= \frac{1}{\pi}\int_{-\infty}^\infty dk\,\ln\left(1+\Lambda e^{-k^2}\right),
\end{equation}
where
\begin{equation*}
\label{Lambda_SSEP}
\Lambda \!=\! n_-(e^\lambda -1) \! +\!  n_+(e^{-\lambda} -1) \! + \! n_- n_+(e^\lambda -1)\,(e^{-\lambda} -1).
\end{equation*}
In this paper we only deal with the \emph{deterministic} (also called \emph{quenched}) setting, when no density fluctuations are allowed at $t=0$. Still, in the special case of $n_{-}= n_{+} =1/2$,
one can obtain $\mu_{\text{quenched}}$ from Eq.~(\ref{annealed}). This is because DG proved,
in the framework of MFT formalism that, in this special case, $\mu_{\text{quenched}}$ is simply related to $\mu_{\text{annealed}}$:
\begin{equation}
\mu_{\text{quenched}}(\lambda, 1/2, 1/2) =\frac{1}{\sqrt{2}} \, \mu_{\text{annealed}}(\lambda, 1/2,1/2)
\label{relation}
\end{equation}
for any $\lambda$  \cite{DG2009b}.  Our inviscid theory is only valid for large currents, $\lambda\gg 1$.  Calculating the $\lambda \gg 1$ asymptotic of Eq.~(\ref{annealed}) (which is actually independent of $n_-$ and $n_+$) and using Eq.~(\ref{relation}), we obtain
\begin{equation*}
   \mu_{\text{quenched}}(\lambda \gg 1,1/2,1/2) \simeq \frac{2\sqrt{2}}{3\pi}\,\lambda^{3/2}.
\end{equation*}
Now we can determine $s(j,n_-,n_+)$ from Eq.~(\ref{reverse}).
The maximum is achieved at $\lambda=\pi^2j^2/6$, and so $s=\pi^2j^3/6$ which yields
$f(1/2,1/2)=\pi^2/6$ as expected.


\begin{thebibliography} {99}
\bibitem{varadhan2}  Recently the $\sim J^3/T$ decay was proved
for the SSEP for more general initial density profiles \cite{varadhan}.
\bibitem{regularization} The presence of a point-like void
at $t=0$ in the inviscid
Eqs. (\ref{qp:eq}) implies a
rapid formation of a deep and narrow density minimum in
the complete MFT equations (\ref{q:eqfull}) and (\ref{p:eqfull}).
Other discontinuities, exhibited by the inviscid solutions,
will be also regularized by diffusion, except for the true singularity
$\partial_x p(x,1) =\lambda \,\delta(x)$ imposed by
Eq.~(\ref{p_step}).
\end{thebibliography}
\end{document}